\documentclass[preprint,prl,aps,amsmath,amssymb,color,superscriptaddress]{revtex4}
\usepackage{stmaryrd}
\usepackage{amsmath}
\usepackage{amssymb}
\usepackage{graphicx}
\usepackage{dcolumn}
\usepackage{bm}
\usepackage{tabularx}
\usepackage{color}
\usepackage{booktabs}
\usepackage{multirow}
\begin{document}
\begin{titlepage}
\title{Identifying topological excitonic insulators via bulk-edge correspondence}

\author{Hongwei Qu}
\affiliation{Key Lab of advanced optoelectronic quantum architecture and measurement (MOE), and Advanced Research Institute of Multidisciplinary Science, Beijing Institute of Technology, Beijing 100081, China}

\author{Zeying Zhang}
\affiliation {College of Mathematics and Physics, Beijing University of Chemical Technology, Beijing 100029, China}

\author{Yuanchang Li}
\email{yuancli@bit.edu.cn}
\affiliation{Key Lab of advanced optoelectronic quantum architecture and measurement (MOE), and Advanced Research Institute of Multidisciplinary Science, Beijing Institute of Technology, Beijing 100081, China}
		
\date{\today}

\begin{abstract}
In this work, we show that the unique bulk-edge correspondence can serve as a fingerprint for identifying topological excitonic insulators, with the LiFe$X$ ($X$ = S, Se, and Te) family as a prototype. First-principles Bethe-Salpeter equation calculations reveal excitonic instabilities in these spin-orbit coupling quantum anomalous Hall insulators. Effective Hamiltonian analyses indicate that spontaneous exciton condensation does not disrupt the gapless edge state but reconstructs the bulk-gap to be almost independent of the spin-orbit coupling strength. This change in the bulk-edge correspondence can be experimentally inspected by angle-resolved photoemission spectroscopy or electron compressibility measurements, providing observational evidence for the identification of topological excitonic insulators.
\end{abstract}

\maketitle
\draft
\vspace{2mm}
\end{titlepage}

\textbf{I. Introduction}

Excitonic insulator (EI) is a strongly correlated semiconductor that harbors spontaneously generated and condensed excitons (electron-hole pairs bound by Coulomb interactions)\cite{Mott,Knox,Kohn,Keldysh,Halperin}. It is a macroscopic quantum system, akin to a superconductor, that essentially stems from many-body interactions among electrons. Instead of the excited states conventionally produced under energy injection, excitons now constitute the ground state of the system. This leads to a scarcity of suitable materials and difficulties in experimental identification\cite{Kogar,Jiangtri,Jianghf}. Despite being pursued since its theoretical conception in the 1960s, there is currently no recognized EI. Thanks to recent advancements in computational science, a number of potential EIs have been predicted, and there is an urgent need to find reliable experimental methods for conclusive identification\cite{usEI,usCPB,WuN,Kaneko,Moon}.

By definition, EIs are simply crystals where the exciton binding energy ($E_b$) exceeds the single-electron gap ($E_g$) at 0 K\cite{Halperin,usEI}. Going to measure the $E_b$ of an EI candidate means that spontaneous excitons are already known to occur in that system. So, verifying EIs by their definition clearly falls into a paradox: one wants to prove an intrinsic material as an EI, but one needs to know explicitly beforehand that there are spontaneous excitons in that material. Contemporary verification of EIs utilizes phenomena derived from the phase transition induced by the spontaneous condensation of excitons. Typically, the excitonic transition leads to abrupt changes in crystal structure, frontier states, gap size, or optical properties that provide identifying signals\cite{Bucher,Cercellier,Wakisaka 2009,Durr}. However, none of these serve as a definitive ``fingerprint", and their presence only indicates the possibility of an EI. To achieve conclusive identification, all other competing mechanisms must be excluded, which is what makes EI identification so challenging. For example, 1$T$-TiSe$_2$ and Ta$_2$NiSe$_5$ are two highly interesting EI candidates. Since phase transitions are accompanied by structural distortions, there has been a debate on whether the driving force is the EI or the Jahn-Teller mechanism\cite{Kogar,Mazza,Guo}. To circumvent this interference, researchers have turned to direct-gap semiconductors, and several EIs without structural distortions have been predicted\cite{usEI}. However, in practice, it is still necessary to distinguish between single-electron and many-body gaps and exclude other possibilities such as Mott and disorder mechanisms\cite{LiuJ 2021,Varsano1}. Although theoretical calculations can in principle help elucidate the gap nature, on the one hand, the EI theory involving quantum many-body problems is not well-established, and on the other hand, it is not realistic to rule out possible competing mechanisms one by one. Some researchers have attempted to probe EIs via superfluidity\cite{LiuXM}, yet such transport measurements are extremely difficult due to the charge-neutral nature of excitons. In fact, there is still controversy in the theoretical community about the existence of superfluidity in EIs\cite{Halperin,Mazza}.

The interplay between strong electronic correlations and band topology opens up new research avenues and provides solutions to long-standing puzzles in condensed matter physics. The realization of the quantum anomalous Hall (QAH) effect is a successful example\cite{Chang}. Topological EI, which combines topological edge states and spontaneous exciton condensation of the bulk, has attracted intense attention from both the experimental and theoretical communities\cite{Durr,Daniele,Wusanfeng,Cobden,DongTEI,LiXZ}. Incorporating topology brings two additional advantages for identifying EI: i) The presence of nontrivial topology excludes some gap mechanisms and thus inherently circumvents their interference. ii) Topological EIs are not only a special class of EIs, but also a special class of topological insulators. Different classes of topological insulators are distinguished by their unique bulk-edge correspondence\cite{JiZ,OuY}. Therefore, deciphering the bulk-edge correspondence of topological materials will, in principle, allow for the conclusive identification of a topological EI.

In this work, we demonstrate how the bulk-edge correspondence enables us to distinguish between topological EIs and conventional spin-orbit coupling (SOC) topological insulators. We elaborate on this using the LiFe$X$ ($X$ = S, Se, and Te) family, which has been predicted to be SOC QAH insulators\cite{LiYang}. To our knowledge, all current topological EIs involve excitonic instabilities in the quantum spin Hall insulator, and it is unknown whether such instabilities can occur in a QAH insulator. We perform first-principles calculations combined with the Bethe-Salpeter equation (BSE) to reveal excitonic instabilities in LiFeS and LiFeSe but not in LiFeTe. We then construct effective Hamiltonians to show that EI formation does not compromise gapless edge states. However, the bulk-gap reconfigured by exciton condensation becomes almost independent of the SOC strength, which is quite different from the linear growth upon SOC enhancement in SOC topological insulators. This difference in gap-SOC dependence can be evaluated experimentally by angle-resolved photoemission spectroscopy or electron compressibility measurements\cite{Young}, thus providing direct and unambiguous evidence for the identification of topological EIs. Finally, we assess the critical temperature of the EI phase to be over 400 K within the mean-field theory, which contributes to the operating temperature of the relevant QAH devices.

\begin{figure}[htb]
	\includegraphics[width=0.8\columnwidth]{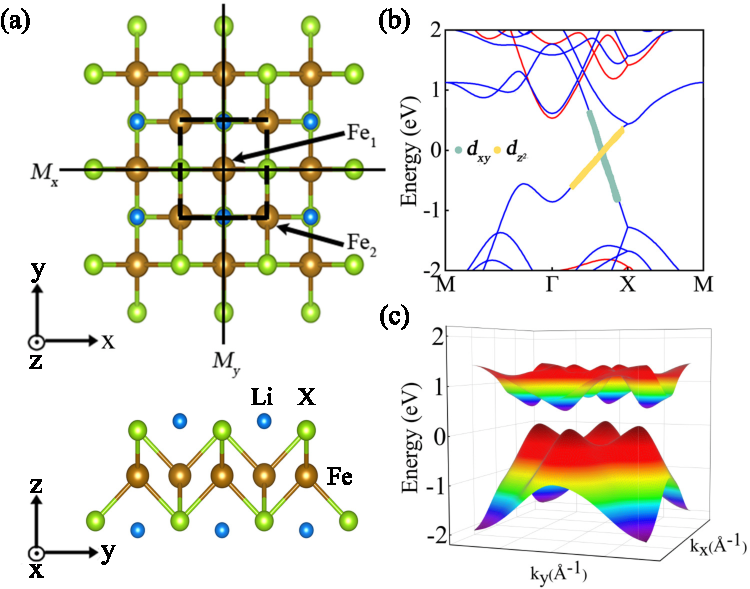}
	\caption{\label{fig:fig1} (a) Top and side views of the monolayer LiFe$X$ structure, which contains an out-of-plane Li-$X$-Fe-$X$-Li quintuple layer and an in-plane tetragonal lattice. The unit cell (black dashed rectangle) has two sets of Li, $X$ and Fe atoms. The $M_x$ and $M_y$ lines represent the two mirror symmetries. (b) Spin-resolved band structure of LiFeSe without considering the SOC, as well as the orbital-projection of the linear Dirac-cone. Red and blue lines denote spin-majority and spin-minority, respectively. (c) Three-dimensional Dirac-cone band structure of LiFeSe in the entire Brillouin zone with the SOC included. In (b) and (c), the Fermi levels are set to zero.}
\end{figure}

\vspace{0.3cm}
\textbf{II. Methodology and models}
\vspace{0.3cm}

Density functional theory (DFT) calculations were performed within the Perdew-Burke-Ernzerhof (PBE) exchange-correlation functional\cite{PBE} and the Heyd-Scuseria-Ernzerhof (HSE) hybrid functional\cite{hse} using the VASP code\cite{vasp}. Electron-ion interaction was described by the projector augmented wave method\cite{PAW,PAW2} with an energy cutoff of 350 eV. A vacuum layer of 20 \AA\ was used to minimize spurious interactions between two neighboring images. A $15 \times 15 \times 1$ $k$-grid was used to find the geometric, electronic and magnetic ground state on the single-electron level. The excitonic properties were obtained by solving the BSE using the YAMBO code\cite{Yambo2} with the single-electron band produced by the QUANTUM ESPRESSO package\cite{QE}. A fine $30\times30\times1$ $k$-grid, 300 bands and 10 Ry cutoff were used to calculate the dielectric function matrix (see Supplemental Materials\cite{SI} for convergence studies). Top four valence and bottom six conduction bands were included to build the BSE Hamiltonian. Given the computational cost, the BSE was solved on top of PBE band but with $E_g$ corrected to the HSE value using a scissor operator for both the response function and diagonal part of the BSE kernel, which has been applied to study excitonic instabilities in low-dimensional materials\cite{Varsano1,usEI,Jianghf,Jiangtri}.

\vspace{0.3cm}
\textbf{III. Results and discussion}
\vspace{0.3cm}

\emph{First-principles results--}The LiFe$X$ family exhibits a similar crystal structure, as depicted in Fig. 1(a). It is composed of a Li-$X$-Fe-$X$-Li quintuple layer with a tetragonal lattice in the $P$4/$nmm$ space group, forming a crystallographic monolayer containing two sets of Li, $X$, and Fe atoms. DFT and HSE calculations neglecting SOC both indicate that all three are 100\% spin-polarized Dirac half-metals [see Fig. 1(b) and Fig. S1 of the Supplemental Materials\cite{SI}] with an integer magnetic moment of 6 $\mu_{\rm B}$. It is the particular Fe 3$d$ occupation and strong kinetic exchange that stabilize LiFe$X$ an ultra-stable ferromagnetic ground state\cite{LiYang}. There is a gap separated by Fe-$d$ and $X$-$s$ states for spin-majority while a linear Dirac-cone formed by Fe $d_{xy}$ and $d_{z^2}$ orbitals appears for spin-minority. The degeneracy at the Dirac points originates from the mirror symmetry of $M_x$/$M_y$ as shown in Fig. 1(a). When SOC is included, LiFe$X$ displays a gap that depends on the easy magnetization axis. An out-of-plane easy axis breaks the $M_x$/$M_y$ symmetry, enabling $d_{xy}$ and $d_{z^2}$ hybridization to generate a gap. In contrast, an in-plane easy axis does not gap the Dirac-cone. Our HSE calculations reveal that LiFeS, LiFeSe, and LiFeTe all have out-of-plane easy axes that open gaps of 0.22, 0.31, and 0.57 eV, as shown in Fig. 1(c) and Fig. S1\cite{SI}. Given that Dirac half-metal is a natural avenue toward the QAH effect\cite{li2005}, we conduct Berry curvature calculations and find that each gapped Dirac-cone contributes a quantized Berry phase of $\pi$. Because there is a total of 4 such valleys, the QAH conductance is $\sigma_{xy}$ = 2$e^2/h$, corresponding to a Chern number $C$ = 2. So, all three LiFe$X$ are SOC QAH insulators, consistent with previous reports\cite{LiYang}.

\begin{figure}[htb]
	\includegraphics[width=1.0\columnwidth]{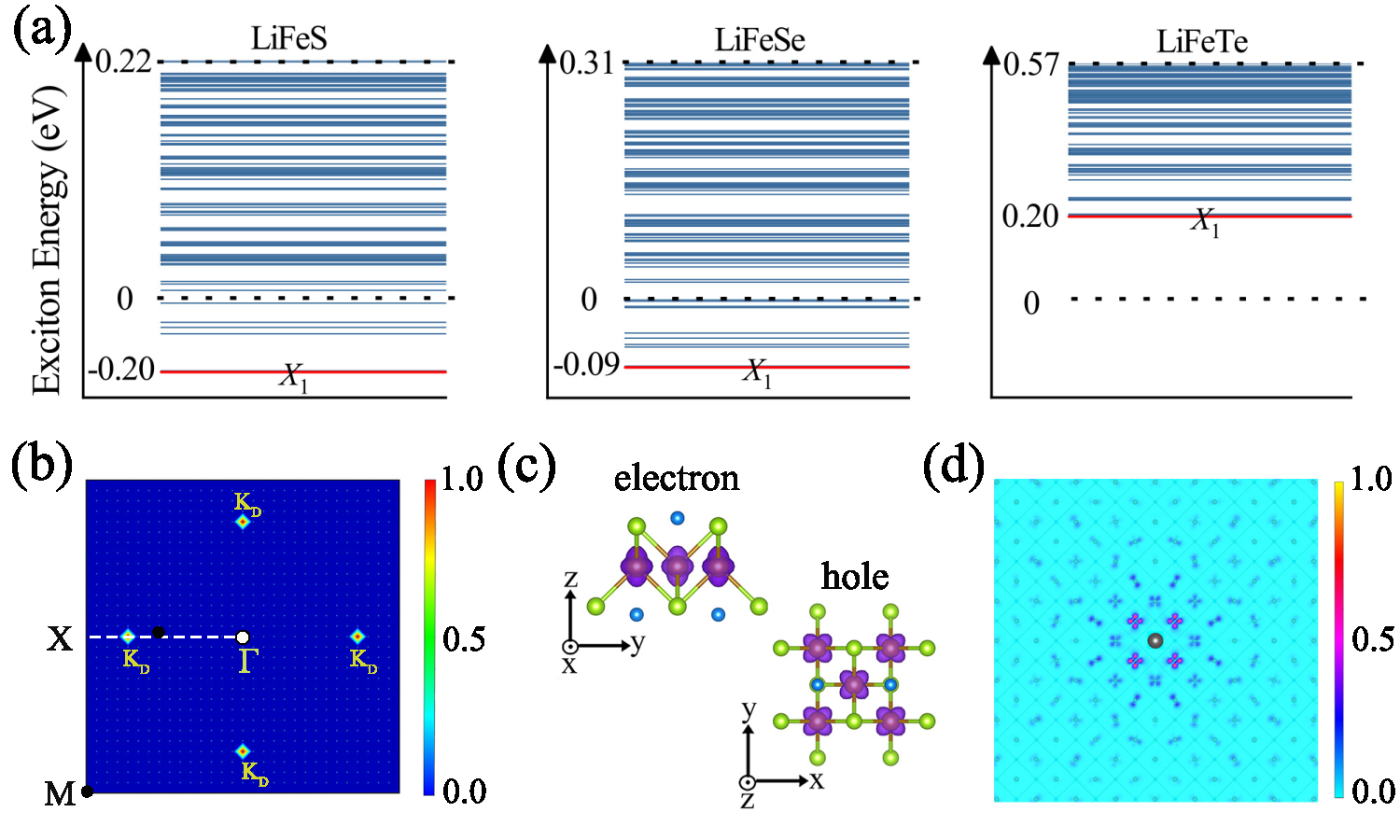}
	\caption{\label{fig:fig2} (a) Exciton formation energy spectra of LiFeS, LiFeSe and LiFeTe obtained from solving the BSE. Each horizontal line represents an exciton state, and the one with the lowest energy is labeled as the $X_1$-exciton. An excitonic instability occurs if the $X_1$-exciton has a negative energy. All shown here are zero-momentum excitons, since our exciton dispersion calculations indicate that the ground-state exciton has \textbf{q} = 0 (see Fig. S3\cite{SI}). (b) Wavefunction modulus of the $X_1$-exciton in the reciprocal space for LiFeSe. (c) Plots of decomposed charge density for electrons and holes that make up the $X_1$-excitons in LiFeSe with an isosurface of 0.1 e/$\rm {\AA}^3$. (d) Wavefunction modulus of the $X_1$-exciton in the real space for LiFeSe, with the hole fixed at the center (black dot). In (b) and (d), the maximum modulus has been renormalized to unity.}
\end{figure}

The Dirac-cone band, monolayer structure, and the predominance of $d$ contributions near the Fermi energy all suggest a weak Coulomb screening in LiFe$X$. This naturally gives rise to significant excitonic effects. We have thus solved the BSE for the low-energy excitation spectrum as shown in Fig. 2(a). It can be seen that the lowest $X_1$-excitons have formation energies ($E_t$) of $-$0.20 and $-$0.09 eV in LiFeS and LiFeSe, respectively, and 0.20 eV in LiFeTe. A negative $E_t$ indicates excitonic instability, meaning that the $X_1$-excitons form spontaneously without the need of energy input. These excitons may reconfigure the ground state at low temperatures, such as into an EI. Hence, LiFeS and LiFeSe possess fundamentally different bulk insulating properties compared to LiFeTe.

Taking LiFeSe as an example, Fig. 2(b) plots the wavefunction of the $X_1$-exciton in reciprocal space. It is almost completely distributed around the four gapped Dirac points, which is characteristic of the Wannier-Mott exciton. Thus, the $X_1$-exciton has a well-defined $E_b$ equal to the difference between the corresponding $E_g$ and $E_t$, i.e., $E_b$ = $E_g - E_t$\cite{Guo}. For LiFeS, LiFeSe, and LiFeTe, the resulting values are 0.42, 0.40, and 0.37 eV, respectively. All $E_b$ are around 0.4 eV, partly because the difference in $E_g$ is not significant, and more likely because the unique nonlocal screening in the monolayer is less sensitive to the constituent elements\cite{Jianglinear,usEI}. On the other hand, first-principles calculations cannot determine the exact value of $E_g$. Typically, different methods yield different results. Nevertheless, the insensitivity of $E_b$ to $E_g$ makes it possible to estimate $E_g \approx$ 0.4 eV as the critical threshold for the occurrence of excitonic instability.

Figure 2(c) shows the charge densities of the electron and hole that make up the $X_1$-exciton. The electron displays a $d_{z^2}$ feature while the hole displays a $d_{xy}$ feature. Being highly localized in the reciprocal space, the $X_1$-exciton extends over a large distance in real space as shown in Fig. 2(d). At the same time, however, the modulus of the electron wavefunction decays rapidly away from the hole, with the intensity mainly concentrated on the neighbouring Fe. In the single-electron picture, two neighbouring Fe atoms are identical. In the excitonic phase, the spontaneous production of $X_1$-excitons causes one of the two Fe atoms to behave like a hole while the other behaves like an electron. As a result, the spatial inversion symmetry is spontaneously broken\cite{Jianghf,DongTEI,Varsano1}.

\emph{Effective Hamiltonians--}Next, we explore the properties of EI ground states. We start with the impact of excitonic phase transitions on topological edge states. Based on the aforementioned first-principles BSE results, we construct an effective Hamiltonian in the basis \{$|$Fe$_1$, $d_{xy}$$>$, $|$Fe$_1$, $d_{z^2}$$>$, $|$Fe$_2$, $d_{xy}$$>$, $|$Fe$_2$, $d_{z^2}$$>$\}
\begin{equation}\label{(1)}
H(\vec{k}) = H_0 + H_{\rm {soc}} + H_{\rm {eh}} =
\begin{pmatrix}
\epsilon_1 & 0   & \vec{t}_1 & \vec{t}_2 \\
0   & \epsilon_2 & \vec{t}_2 & \vec{t}_3 \\
\vec{t}_1 & \vec{t}_2 & \epsilon_1 & 0 \\
\vec{t}_2 & \vec{t}_3 & 0   & \epsilon_2
\end{pmatrix}
+
\begin{pmatrix}
0 & -\vec{r} & 0 & 0 \\
\vec{r} & 0 & 0 & 0 \\
0 & 0 & 0 & -\vec{r} \\
0 & 0 & \vec{r} & 0
\end{pmatrix}
+
\begin{pmatrix}
0 & 0 & 0 & \vec{\Lambda}_1 \\
0 & 0 & \vec{\Lambda}_2 & 0 \\
0 & \vec{\Lambda}_2 & 0 & 0 \\
\vec{\Lambda}_1 & 0 & 0 & 0
\end{pmatrix}.
\end{equation}
Here, $H_0$ and $H_{\rm {soc}}$ are formulated in the tight-binding approximation using the MagneticTB package\cite{MagneticTB} developed by one of the authors. The parameters $\epsilon_{1(2)}$, $\vec{t}_{1(2,3)}$ and $\vec{r}$ represent the orbital energy, hopping and SOC parameters, respectively. For more details, see the Supplemental Materials\cite{SI}. $H_{\rm {eh}}$ describes the effect of $X_1$-excitons, which stems from two facts of the first-principles result: (i) It involves the pairing between $d_{xy}$ and $d_{z^2}$ of Fe$_1$ and Fe$_2$. (ii) The equivalence of Fe$_1$ and Fe$_2$ is broken, so $\vec{\Lambda}_1$ and $\vec{\Lambda}_2$ must be different. For simplicity, let $\vec{\Lambda}_1$ = 0.

\begin{figure}[htb]
	\includegraphics[width=0.8\columnwidth]{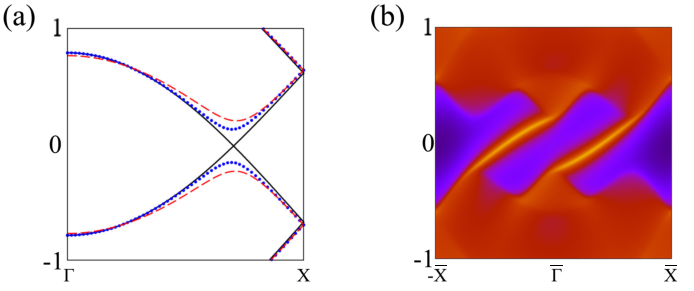}
	\caption{\label{fig:fig1} (a) Band structures derived from $H$($\vec{k}$) under different scenarios, namely, $H_0$ (black lines), $H_0$$+$$H_{\rm {soc}}$ (blue dots), and $H_0$$+$$H_{\rm {soc}}$$+$$H_{\rm {eh}}$ (red dashes). See the Supplementary Material\cite{SI} for more details. (b) Edge modes obtained from $H_0$$+$$H_{\rm {soc}}$$+$$H_{\rm {eh}}$, where there are two gapless edge states (bright yellow lines) connecting the conduction and valence bands. The Fermi levels are set to zero.}
\end{figure}

Figure 3(a) depicts the bands derived from $H$($\vec{k}$) under different scenarios. When only $H_0$ is considered, a Dirac-cone appears (black lines). Adding $H_{\rm {soc}}$ lifts the degeneracy of the Dirac point and creates a gap (blue dots), which is further enlarged by $H_{\rm {eh}}$ (red dashes). These findings are in good agreement with the first-principles results. Subsequently, we compute the edge states using the Hamiltonian (1) with and without $H_{\rm {eh}}$, respectively. The former result is presented in Fig. 3(b). There are two bands connecting the conduction and valence bands, a feature that is the same as the case without $H_{\rm {eh}}$ (see Fig. S4 of Supplemental Materials\cite{SI}). Thus, the excitonic phase transition does not compromise the gapless edge states of the SOC QAH phase. In other words, the EI phase remains topologically nontrivial with $C$ = 2. This finding is of scientific significance. It demonstrates that excitonic instability in SOC topological insulators leads to topological EIs without causing a topological phase transition, validating previous speculations\cite{DongTEI,Wu2015}.

We proceed to study the bulk-gap of the EI phase using a simple two-band effective Hamiltonian\cite{Kaneko,note}
\begin{align}
	\hat{H} & =\sum_{\vec{k}}[\varepsilon_{a}(\vec{k})-\mu]a_{\vec{k}}^+a_{\vec{k}}+\sum_{\vec{k}}[\varepsilon_{b}(\vec{k})-\mu]b_{\vec{k}}^+b_{\vec{k}}\nonumber\\
&     +\frac{1}{2S}\sum_{\vec{k'},\vec{k},\vec{q}}[V_{aa}(\vec{q})a_{\vec{k}+\vec{q}}^+a_{\vec{k'}-\vec{q}}^+a_{\vec{k'}}a_{\vec{k}}+V_{bb}(\vec{q})b_{\vec{k}+\vec{q}}^+b_{\vec{k'}-\vec{q}}^+b_{\vec{k'}}b_{\vec{k}}-2V_{ab}(\vec{q})a_{\vec{k}+\vec{q}}^+b_{\vec{k'}-\vec{q}}^+b_{\vec{k'}}a_{\vec{k}}],
\end{align}
where $\mu$ and $S$ are the chemical potential and the in-plane area. $a_{\vec{k}}$ ($a_{\vec{k}}^{\dag}$) and $b_{\vec{k}}$ ($b_{\vec{k}}^{\dag}$) are the destruction (creation) operators of electron and hole. $\varepsilon_{a}(\vec{k})=\frac{\hbar^2 \vec{k}^2}{2m_{\rm e}}+\frac{E_{g}}{2}$ and $\varepsilon_{b}(\vec{k})=-\frac{\hbar^2 \vec{k}^2}{2m_{\rm h}}-\frac{E_{g}}{2}$ denote the conduction and valence band within single-electron picture\cite{Jianglinear,Jiangtri}, with electron (hole) effective mass $m_{\rm e}$ ($m_{\rm h}$) and $E_g$ fitted from our first-principles calculations including the SOC, i.e., Fig. 1(c). $V(\vec{q}$) denotes the many-body interactions, i.e., $V(\vec{q})=V_{aa}(\vec{q})=V_{bb}(\vec{q})=V_{ab}(\vec{q})=\frac{2\pi}{(1+2\pi\alpha_{2D}|q|)|q|}$, with two-dimensional polarizability $\alpha_{2D}$ obtained from first-principles calculations. Using the Hartree-Fock approximation and $\frac{\varepsilon_a(\vec{k})+\varepsilon_b(\vec{k})}{2}$ = 0\cite{Xue}, we derive the coupled equations
\begin{equation}
	\begin{array}{ll}
		~~~~~~~~~~~~~~~~~~	\Delta(\vec{k}) = \frac{1}{2S} \sum_{\vec{k'}} V(\vec{k} - \vec{k'}) \frac{\Delta(\vec{k'})}{E(\vec{k'})},\\
		~~~~~~~~~~~~~~~~~~  \xi(\vec{k}) = \frac{\varepsilon_{a}(\vec{k}) - \varepsilon_{b}(\vec{k})}{2}-\frac{1}{2S}\sum\limits_{\vec{k'}}V({\vec{k}-\vec{k'}})(1-\frac{\xi(\vec{k'})}{E(\vec{k'})}),\\
		~~~~~~~~~~~~~~~~~~	E(\vec{k}) = \sqrt{\xi^2(\vec{k}) + \Delta^2(\vec{k})}.
	\end{array}
\end{equation}
Here $\Delta(\vec{k})$ is defined as the order parameter of the EI phase and $\xi(\vec{k})$ is an auxiliary quantity. $E(\vec{k})$ is the reformulated band spectrum and twice its minimum value gives the bulk-gap of the EI phase\cite{Kaneko}.

\begin{figure}[htb]
	\includegraphics[width=0.8\columnwidth]{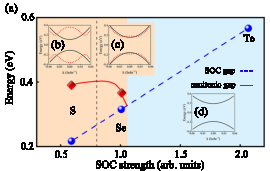}
	\caption{\label{fig:fig1} (a) Bulk-gaps of the LiFe$X$ as a function of SOC strength in the SOC and excitonic QAH phases, respectively.
The SOC bulk-gaps are obtained by first-principles (Blue balls), while the excitonic ones (Red diamonds) are obtained by self-consistently solving for $E(\vec{k})$ in Eqs. (3). Dependence curves are established by linearly interpolating the first-principles results of LiFeS, LiFeSe, and LiFeTe in order to simulate the effect of possible $S$-group atom alloying\cite{XuSY}. See the Supplemental Materials\cite{SI} for more details. The blue and orange regions indicate that the system ground-state is in the SOC and excitonic QAH phase, respectively. Comparison between the band spectra of (b) LiFeS and (c) LiFeSe in the SOC (black solid) and excitonic (red dotted) QAH phases. The black dashed vertical line distinguishes the left and right excitonic QAH regions, in which the exciton reformulated band spectra have a Mexican-hat and a flattened shape, respectively. (d) Band-edge spectrum of LiFeTe in the SOC QAH phase.}
\end{figure}

Figure 4(a) summarizes the bulk-gap of LiFe$X$ in the SOC and excitonic QAH phases, respectively. Note that LiFeTe does not have excitonic instability. In fact, by utilizing the mutual solubility between $S$-group elements to tune the SOC strength of the system\cite{XuSY}, it is possible to obtain a richer phase diagram experimentally, rather than being limited to just three points. To this end, we have further established a gap-SOC curve by linearly interpolating the first-principles results of LiFeS, LiFeSe, and LiFeTe. See the Supplemental Materials\cite{SI} for details. It can be seen that the bulk-gaps of the two phases exhibit very different trends. When in the SOC QAH phase, the bulk-gap increases linearly as the SOC increases from S to Se to Te. When in the excitonic QAH phase, the bulk-gap varies much less with SOC. When LiFe$X$ are all in their own ground states, their bulk-gaps have a LiFeTe $>$ LiFeS $>$ LiFeSe ordering, which is distinctly different from that in the SOC QAH phase, where LiFeTe $>$ LiFeSe $>$ LiFeS. The difference is also reflected in the variation trend. While replacing Se with Te in LiFeSe increases the SOC, the excitonic bulk-gap decreases until the excitonic instability disappears. This qualitative difference thus allows a clear experimental determination of whether LiFe$X$ is in the topological EI state.

Figures 4(b)-(d) shows a comparison of the band spectra, respectively, in the SOC and excitonic QAH phases. In the SOC phase, LiFe$X$ all have similar parabolic shapes. In the excitonic phase, LiFeS reformulates to a Mexican-hat shape while LiFeSe reformulates to a flatter shape. Band reshaping provides evidence for identifying the excitonic QAH phase using angle-resolved photoemission spectroscopy, which also has an impact on the transition temperature (see Supplemental Materials\cite{SI} for details).

Exciton reconstructing the band spectrum necessarily produces a distinct bulk-edge correspondence from the SOC topological phase. As bulk-edge correspondence is the hallmark that distinguishes different classes of topological matters\cite{JiZ,OuY}, identifying topological EIs by bulk-edge correspondence is not only universally effective beyond the LiFe$X$ family, but is also conclusive. In particular, first-principles calculations, which enjoy great success in predicting and confirming topological materials, can contribute to topological EIs as well. When the emergence of gapless edge states locks the system in a topological phase, the bulk-gap may come from either the SOC or the exciton. In the former case, the bulk-gap necessarily increases with increasing SOC strength, so that the heavier the constituent element, the larger the bulk-gap\cite{XuSY}. In the latter case, the bulk-gap is approximated by $E_b$, which depends on the overall screening effect of the system. For low-dimensional materials, the $E_b$ is not sensitive to the constituent elements\cite{Dong2020,DongTEI,LiuJ 2021}. Therefore, topological EIs can be unambiguously identified by modulating the system SOC strength through substituting elements of the same group and then monitoring the gap-SOC dependence. Bulk-gaps that vary significantly and monotonically with SOC are SOC topological insulators, whereas those that do not vary much and have no significant dependence are topological EIs. It should be noted that there is a lack of direct observations of excitons here. In order to completely rule out other possibilities, a confirmation can be made by combining methods such as the exciton compressibility measurement.

\emph{Critical temperature--}Within the excitonic QAH (topological EI) phase, the gapless edge states are protected by the order parameter\cite{Mai}. At this point, the critical temperature $T_{\rm c}$ is determined when the minimum $\Delta(\vec{k})$ is zero. Multiplying Eqs. (3) by the temperature factor ($\frac{1}{e^{\frac{-E(\vec{k})}{k_{\rm {B}}T}}+1}-\frac{1}{e^{\frac{E(\vec{k})}{k_{\rm {B}}T}}+1}$)\cite{Jiangtri}, we can estimate the $T_{\rm c}$ of LiFeS and LiFeSe as 1050 and 400 K, respectively. Such $T_{\rm c}$'s are much lower than the ferromagnetic Curie temperatures $\sim$1700 and 1500 K for LiFeS and LiFeSe\cite{LiYang}, indicating that the excitonic QAH phase is robust to the magnetic state. Hence, QAH devices based on LiFeS and LiFeSe are expected to operate at temperatures above room temperature. More interestingly, the formation of EI leads to a much higher operating temperature for LiFeS than for LiFeSe, which is an opposite trend to that of SOC topological insulators. This not only provides evidence for the EI identification, but also indicates that lighter elements are more favourable for topological EIs, which greatly complements the selection range of SOC topological materials.

Above estimation treats the spontaneously formed excitons as a weakly interacting boson gas. If treated as an ideal boson gas, the $T_{\rm c}$ can be calculated using the statistical formula\cite{NJP}
\begin{equation}
\begin{split}
\label{eq5}
n = \frac{{ mk_{\rm B}T_{\rm c}}}{{2\pi {\hbar^2}}}\sum\limits_{j = 1}^\infty  {\frac{{{{({e^{-|{E_t}|/k_{\rm B}T_{\rm c}}})}^j}}}{{{j}}}},
\end{split}
\end{equation}
where $m$ and $n$ are exciton mass and density, respectively. The former is obtained by fitting first-principles results, while the latter is estimated using Eqs. (3), i.e., $n=\sum\limits_{\vec{k}}\frac{|\Delta(\vec{k})|^2}{S}$\cite{Ataei,Daniele,Xue}. It leads to $n= 1.3\times10^{12}$ and $9.7\times10^{10}$ cm$^{-2}$ for LiFeS and LiFeSe, respectively, which corresponds to $T_{\rm c}^{\rm {sta}}$ = 1126 and 334 K. Compared to the results of Eq. (3), $T_{\rm c}$ increases by 76 K for LiFeS while $T_{\rm c}$ decreases by 66 K for LiFeSe. The small difference manifests the weak inter-exciton interactions.

Electrons and holes pair by direct Coulomb attraction in EIs, unlike in BCS superconductors where electron pairing is mediated by phonons. As a result, the coupling strength in the former is much stronger than in the latter. For example, the order parameter of LiFeS is about 200 meV, which is two orders of magnitude larger than several meV of a typical BCS superconductor with a $T_c$ of 1$\sim$10 K\cite{BCS1,BCS2}. Thus, it is not surprising that a similar mean-field theory predicts $T_c$ = 1050 K. Whilst, for monolayers, the mean-field theory may overestimate the superfluid transition temperature that is determined by the Kosterlitz-Thouless (KT) temperature. Using $T_{KT} = \frac{n\pi\hbar^2}{2mk_B}$\cite{BKT1,BKT2}, we have $T_{KT}$ = 64.3 and 3.2 K for LiFeS and LiFeSe, respectively.

\vspace{0.3cm}
\textbf{IV. Conclusion}
\vspace{0.3cm}

In summary, our first-principles BSE combined with model Hamiltonian calculations reveal that the monolayer LiFeS and LiFeSe are excitonic QAH insulators, while the LiFeTe is a SOC QAH insulator. The excitonic topological phase exhibits unique bulk-edge correspondence that is quite different from that of the conventional SOC topological phase, in particular the bulk-gap dependence on the SOC, which provides a hallmark for unambiguous identification. Although the findings are drawn from the LiFe$X$ family, the bulk-edge correspondence principle is certainly applicable to identifying any topological EI, as different topological matters are essentially distinguished by their bulk-edge correspondence. Our work not only makes progress toward solving a long-standing challenge of unambiguously identifying EIs, but also extends the understanding of topological insulators and offers new perspectives on enhancing the operational temperature of QAH devices.

\vspace{0.3cm}
\textbf{Acknowledgments}
\vspace{0.3cm}

Y.L. thanks Z. Liu and X. M. Zhang for a useful discussion. This work was supported by the Ministry of Science and Technology of China (Grant Nos. 2023YFA1406400 and 2020YFA0308800), the National Natural Science Foundation of China (Grant No. 12074034), and the Fundamental Research Funds for the Central Universities (No. ZY2418).

\vspace{0.3cm}
\textbf{Data availability statement}
\vspace{0.3cm}

The data that support the findings of this study are available upon reasonable request from the authors.


\begin{thebibliography}{90}
\bibitem{note} The oscillator strength (see Ref. \onlinecite{Halperin} for more details) is not considered here because our calculations based on the eigenfunctions of Hamiltonian (1) show that it only causes a change in the bulk-gap of the excitonic QAH phase on the order of meV.
\end{thebibliography}
\end{document}